\documentstyle[12pt]{article}
\input epsf

\begin{document}
\begin{titlepage}
\begin{center}
July 31, 1997\hfill    UND-HEP-97-US02 \\
               \hfill    hep-ph/9707546
\vskip .2in
{\large \bf Low-Energy Signals for a Minimal Gauge-Mediated Model}
\vskip .3in

\vskip .3in
Emidio Gabrielli$^*$\\[.03in]
\vskip 10pt
Uri Sarid$^\dag$\\[.03in]
\vskip 15pt
{\em Department of Physics\\
     University of Notre Dame\\
     Notre Dame, IN 46556\\
     USA}
\end{center}
\vskip 10pt
\begin{abstract}
\medskip
The inclusive branching ratio $B \to X_s\gamma$ and the anomalous magnetic moment $g_\mu-2$ of the muon are accurately calculated within a minimal gauge-mediated SUSY-breaking model which naturally generates a large $\tan\beta$. The predictions are in somewhat better agreement with current experiments, and new data will soon critically test these predictions. Predictions for $B \to X_s\ell^+\ell^-$ branching ratios and asymmetries, to be tested at future colliders, are also presented.
\end{abstract}
\end{titlepage}

Extensions of the standard model (SM) based on supersymmetry (SUSY) whose breaking is conveyed to the observable sector via the usual gauge interactions have received much attention lately \cite{ref:GM}, for several reasons: gauge-mediation models usually predict a rich and distinctive phenomenology based on very few parameters; they naturally suppress flavor-changing neutral current processes; and their SUSY-breaking scale is low enough to avoid quantum-gravity effects and perhaps even to be accessible to future experiments.
  
In this work we examine several low-energy predictions of the minimal gauge-mediated SUSY-breaking model with a single SU(5)-fundamental representation of messenger fields. Neither the $\mu$ term nor its scalar analogue $\mu B$ are generated by gauge mediation; we will assume that $\mu$ eventually arises in such a way that $\mu B$ remains essentially zero at the messenger scale $M_M$, and that the other Higgs parameters are not altered, as explained in Ref.~\cite{ref:rs}. Such a model, which we will refer to as the MGM, generates a large Higgs VEV hierarchy: ${\tan\beta} \equiv v_U/v_D = m_A^2/(\mu B)$ where $m_A$ is the mass of the pseudoscalar Higgs. The superpartner-scale $B$ term is small because there is little room for RG evolution {\it and} because the gauge- and Yukawa-coupling contributions to this evolution tend to cancel. Since ${\tan\beta}$ is {\it naturally} large, the MGM is preferable to the gravity-mediation large ${\tan\beta}$ models; with $B$ (and $A$) induced only radiatively, the SUSY CP problem is solved; ${\tan\beta}$ and $\mu$ are predicted; and the large value of ${\tan\beta}$ dramatically enhances certain processes, as we show below. We will extensively use the results of Ref.~\cite{ref:rs} which studied this model. 

We begin by sharpening the predictions first presented in Ref.~\cite{ref:rs} for the $b$-quark decay $b \to s \gamma$. We then complement them with an even sharper prediction for the muon anomalous magnetic moment $g_\mu - 2$. The correlated prediction constitutes a striking signal for the MGM, and at present is actually in better agreement with experiment than the SM. We also predict branching ratios and asymmetries for $b\to s\ell^+\ell^-$, which are correlated with $b \to s \gamma$ through their common dominant dependence on a single Wilson coefficient in the effective Hamiltonian. Experiments now in progress will soon yield much more precise tests of the $b\to s\gamma$ and $g_\mu-2$ signals, and within a few years the dilepton predictions will also confront experiment.

Before studying the signals, we define our signs. Whenever $A$ and $B$ are entirely radiatively generated, there is only a single physical and predicted sign associated with the various mass parameters. (We will not be concerned with any phases in the quark mixing matrices.) Two steps are required to determine it after fixing {\it any} definite Lagrangian sign convention: first, the scalar potential is minimized, yielding $1/{\tan\beta} = \sigma_1\,\mu B/ m_A^2$ ; then the 1-loop diagram generating $A_t$ from the gluino mass $M_3$, and the 1-loop diagrams generating $\mu B$ from $A_t$ or from the wino mass $M_2$, are calculated to find $\mu B \propto \sigma_2 \mu M_{1/2}$. Here $\sigma_i = \pm1$. The second sign, $\sigma_2$, is determined by a competition between $A_t$ and $M_2$ in generating $B$: in the MGM, the former wins. Thus the MGM predicts ${\rm sgn}(\mu M_{1/2} {\tan\beta}) = \sigma_1 \sigma_2$ within the given convention. In this way we can correlate our findings with others found in the literature.  In the following, $\mu$, ${\tan\beta}$, and the gaugino masses will all be positive.  The convention-independent, physical signs of our predictions will be the relative signs of the various amplitudes for $b\to s\gamma$ (encapsulated in $R_7$) and for $b\to s\ell^+\ell^-$, and the positive contribution to $g_\mu-2$.

A much more detailed presentation than can be fit into this Letter will be given in Ref.~\cite{ref:gslong}.  Below we use the following parameter values: $m_t^{\rm pole} \simeq m_t^{\rm \overline{MS}}(m_Z) \simeq 174\,\rm GeV$, $m_b^{\rm pole} = 4.8\,\rm GeV$, $m_c^{\rm pole} = 1.3\,\rm GeV$, $\mu_b = m_b$ (the low renormalization scale at which $b$ processes will be computed), $\alpha_s(m_Z) = 0.118$, $\alpha_{\rm em}^{-1}(m_Z) = 128.$, and $\sin^2\theta_W = 0.23$.

We first present the amplitudes contributing to the Wilson coefficient $C_7(m_W)$ of the main operator, $Q_7 \propto \overline s \sigma^{\mu\nu} (1 + \gamma_5) b F_{\mu\nu}$, responsible for radiative quark decay $b\to s \gamma$. (Related studies appeared in Refs.~\cite{ref:bsg}.) For large ${\tan\beta}$ the SUSY diagrams are dominated by chargino loops, and to a much lesser degree by the gluino loop. Unlike in Ref.~\cite{ref:rs}, we diagonalize the chargino mass matrix exactly, since we are interested in light charginos. The other mass-insertion approximations we use are quite sufficient. Up to an overall normalization, we find: 
\begin{eqnarray}
{\cal A}_{\rm SM} &=& \frac32 V_{ts} x_{tW} \left[\frac23 F_1(x_{tW}) + F_2(x_{tW})\right]  \label{eq:ASM}\\
{\cal A}_{\rm H^-} &=& \frac12 V_{ts} r_b x_{tH} \left[\frac23 F_3(x_{tH}) + F_4(x_{tH})\right] \label{eq:A2H}\\
{\cal A}_{\rm \tilde h^-} &=& \frac12 V_{ts} r_b{\tan\beta} {\mu m_t m_{\tilde t_L \tilde t_R}^2 \over m_{\tilde t_L}^2 m_{\tilde t_R}^2} \left[M_2^2{F(x_{\tilde h_1 \tilde t},x_{\tilde h_2 \tilde t})\over M_{\tilde h_1^-}^2 - M_{\tilde h_2^-}^2} \right. \nonumber\\
& & \left. \phantom{.} - |m_{\tilde t_L}  m_{\tilde t_R}| {F'(x_{\tilde h_1 \tilde t},x_{\tilde h_2 \tilde t})\over  M_{\tilde h_1^-}^2 - M_{\tilde h_2^-}^2} \right]
\label{eq:AHH}\\
{\cal A}_{\rm \widetilde W \tilde h^-} &=& r_b{\tan\beta} {m_W^2 m_{\tilde t_L \tilde c_L}^2 \over m_{\tilde t_L}^2 m_{\tilde c_L}^2} \mu M_2 {F(x_{\tilde h_1 \tilde q_L},x_{\tilde h_2 \tilde q_L}) \over M_{\tilde h_1^-}^2 - M_{\tilde h_2^-}^2} \label{eq:AWH}\\
{\cal A}_{\rm \tilde g} &=&  \frac89 r_b{\tan\beta} {\alpha_s\over\alpha_W} {m_W^2 m_{\tilde t_L \tilde c_L}^2 \mu M_3 \over m_{\tilde q}^6} F_{\rm gl}(x_{M_3 \tilde q}) \,, \label{eq:AGL}
\end{eqnarray}
where $r_b \equiv 1/(1+{\delta m_b/m_b})$ \cite{ref:rs}. The functions $F_{1,2,3,4}$ were given in Ref.~\cite{ref:bbmr}, while we have defined $F(x_1,x_2) = f(x_1) - f(x_2)$, $F'(x_1,x_2) = x_1 f(x_1) - x_2 f(x_2)$, $f(x) \equiv d(x F_3 + \frac23 x F_4)/dx$, and $F_{\rm gl}(x) \equiv \frac12 d^2(x^2 F_4)/dx^2$. Also, $x_{tW} = m_t^2/m_W^2$; $x_{tH} = m_t^2/m_{H^-}^2$; $x_{\tilde h_i \tilde t} = M_{\tilde h_i^-}^2/|m_{\tilde t_L}  m_{\tilde t_R}|$ where $M_{\tilde h_i^-}$ is the $i$th chargino mass eigenvalue; $x_{\tilde h_i \tilde q_L} = M_{\tilde h_i^-}^2/|m_{\tilde t_L} m_{\tilde c_L}|$ ; and $x_{M_3 \tilde q} = M_3^2/m_{\tilde q}^2$ where $m_{\tilde q}$ is the average squark mass.
The left-right stop mass insertion is $m_{\tilde t_L \tilde t_R}^2 = m_t A_t$ ($>0$), while the left-left stop-scharm mass insertion is  \cite{ref:rs} $m_{\tilde t_L \tilde c_L}^2 \simeq +V_{ts} m_{\tilde q}^2 \lambda_t^2 \ln(M_M/m_{\tilde t})/4\pi^2$. In the MGM, the two chargino amplitudes interfere destructively with the SM, charged-Higgs and gluino amplitudes. They may all be combined into a ratio $R_7$ expressing the fractional amplitude deviation from the SM:
\begin{equation}
R_7 \equiv {{\cal A}_{\rm H^-} + {\cal A}_{\rm \tilde h^-} + {\cal A}_{\rm \widetilde W \tilde h^-} + {\cal A}_{\rm \widetilde g} \over {\cal A}_{\rm SM}}\,.
\end{equation}

The contributions to the related operator $Q_8$, expressed as an amplitude deviation $R_8$ from the SM, are given by the same expressions as above after replacing: $\frac23 F_1 + F_2 \to F_1$, $\frac23 F_3 + F_4 \to F_3$, $f \to d(x F_4)/dx$, and $F_{\rm gl}(x) \to  \frac{9}{16} d^2(3 x^2 F_3 + \frac13 x^2 F_4)/dx^2$. We have included them in the MGM predictions for $b\to s\gamma$, presented below.

From these amplitudes we infer the inclusive $B$ meson branching ratio ${\rm BR}_\gamma \equiv {\rm BR}(B\to X_s \gamma)$ by including the (percent-level) nonperturbative corrections \cite{ref:bsgNP}. We have reevaluated the NLO SM expressions of Refs.~\cite{ref:bsgNLO}, substituting $C_{7,8}^{SM}(m_W) \to (1 + R_{7,8}) C_{7,8}^{SM}(m_W)$ to obtain:
\begin{eqnarray}
10^4\,{\rm BR}^{\rm NLO}_\gamma &=& (3.48\pm 0.31) \left(1 + 0.622 R_7 + 0.090 R_7^2\right.\nonumber\\
&+& \left.0.066 R_8 + 0.019 R_7 R_8 + 0.002 R_8^2\right)
\label{eq:bsgNLO}
\end{eqnarray}
which may be compared to the present experimental findings of CLEO \cite{ref:bsgEXPT}:
\begin{equation}
10^4\,{\rm BR}^{\rm expt}_\gamma = 2.32 \pm 0.67\,,
\label{eq:bsgEXPT}
\end{equation}
and $1.0 < 10^4\,{\rm BR}^{\rm expt}_\gamma < 4.2$ at $95\%$ CL. The SM prediction is higher than the observed branching ratio, although the disagreement is less than two experimental standard deviations.
The central value of our SM branching ratio normalization factor in Eq.~(\ref{eq:bsgNLO}) is in complete agreement with the second paper in Ref.~\cite{ref:bsgNLO}, whose estimate of the overall normalization uncertainties we have adopted here.  But it is also important to consider other uncertainties, in particular in the functional dependence on $R_{7,8}$. Our expression includes the complete SM NLO calculation, and in particular the QCD correction to the SM amplitude at the electroweak scale $C_{7,8}^{\rm SM,(1)}$. But we do not have the corresponding corrections to the other amplitudes (which would be denoted $R_{7,8}^{(1)}$). Also, if bounds on $R_7$ are derived by comparing Eq.~(\ref{eq:bsgNLO}) with data, those bounds depend on $R_8$, albeit weakly; since $R_8$ is model-dependent, fixing it in order to obtain bounds only on $R_7$ introduces some error into those bounds. Considering these and other factors, we estimate that, in addition to the explicit experimental and theoretical uncertainties in Eqs.~(\ref{eq:bsgNLO},\ref{eq:bsgEXPT}), our bounds on $R_7$ carry a further {\it additive} uncertainty $\sim \pm0.10$.

We turn now to the flavor-changing dilepton decays $b\to s\ell^+\ell^-$. The differential decay rate $d^2 \Gamma_{\ell\ell}/dy_+ dy_-$ is conventionally parameterized \cite{ref:bsll} by kinematic factors multiplying Wilson coefficients of various operators defined at some scale $\mu_b = {\cal O}(m_b)$. The kinematic variables are $y_{\pm} = 2 E_{\pm}/m_b$ where $E_{\pm}$ are the lepton energies measured in the $b$ rest frame. We prefer the combinations $\hat s = y_+ + y_- - 1$ and $\hat y = y_+ - y_- = -\hat y_{\rm max} \cos\theta$ where $y_{\rm max} = (1 - \hat s) \sqrt{1 - 4 x_\ell/\hat s}$, $x_\ell = m_\ell^2/m_b^2$ and $\theta$ is the angle between the $b$ and the $\ell^+$ in the $\ell^+\ell^-$ rest frame. We have recomputed these kinematic factors, and will present them explicitly elsewhere \cite{ref:gslong}. 
The branching ratio for a particular $\ell$ is simply proportional to the integral of $d^2 \Gamma_{\ell\ell}/dy_+ dy_-$ over the entire kinematically-allowed range for that $\ell$. The forward-backward or energy asymmetry is proportional to the integral over the range $y_- > y_+$ (i.e., $\cos\theta > 0$) minus the integral over the range $y_- < y_+$ ($\cos\theta < 0$). The kinematic limits are $-\hat y_{\rm max} < \hat y < +\hat y_{\rm max}$ and $4 x_\ell < \hat s < 1$; but to avoid intermediate charmonium resonances and  nonperturbative phenomena near the end point, we exclude the same ranges of $\hat s$ specified in the second reference of \cite{ref:bsll}. 
As usual, we normalize to the semileptonic decay rate $b\to c\,e\overline\nu_e$ in order to eliminate the large dependence on the $b$ quark mass: $\Gamma_{\ell\ell} \to \widehat\Gamma_{\ell\ell} \equiv \Gamma_{\ell\ell}/\Gamma_{b\to c}$, hence
\begin{eqnarray}
{\rm BR}_{\ell\ell} &=&  ({\rm BR}_{b\to c}) \int d\hat s \,2\!\!\int_0^{\hat y_{\rm max}} d\hat y \left({d^2 \widehat\Gamma_{\ell\ell}\over d\hat s d\hat y}\right)_{\rm symm}
\label{eq:BRll} \\
{\rm A}_{\ell\ell}^{\rm FB} = {\rm A}_{\ell\ell}^{\rm E} &\equiv& 
{N(y_- > y_+) - N(y_+ > y_-) \over N(y_- > y_+) + N(y_+ > y_-)} \nonumber\\
&=&
-{1\over\widehat\Gamma_{\ell\ell}} \int d\hat s \,2\!\!\int_0^{\hat y_{\rm max}} d\hat y \left({d^2 \widehat\Gamma_{\ell\ell}\over d\hat s d\hat y}\right)_{\rm antisymm}
\label{eq:All}
\end{eqnarray}
in which the symmetric (antisymmetric) rate must include only the kinematic coefficients symmetric (antisymmetric) in $\hat y$. 

Returning now to the minimal gauge-mediated model, we find that {\it only} the contributions to $R_7$ (rather than box diagrams, etc.) can significantly affect the branching ratios or asymmetries beyond the percent level (see also Ref.~\cite{ref:hewells}), so the deviations in $b\to s\ell^+\ell^-$ are closely correlated to those in $b\to s\gamma$. Numerically, we find:
\begin{eqnarray}
10^7\,{\rm BR}_{ee} &=& 68.5 + 22.4 R_7 + 6.1 R_7^2 \label{eq:BRee} \\
{\rm A}_{ee} &=& (4.52 - 3.01R_7) / (10^7 \,{\rm BR}_{ee}) \label{eq:Aee} \\
10^7\,{\rm BR}_{\mu\mu} &=& 44.64 + 2.46 R_7 + 1.81 R_7^2 \label{eq:BRmumu} \\
{\rm A}_{\mu\mu} &=& (4.68 - 2.90 R_7) / (10^7 \,{\rm BR}_{\mu\mu}) \label{eq:Amumu} \\
10^7\,{\rm BR}_{\tau\tau} &=& 2.013 - 0.201 R_7 + 0.009 R_7^2 \label{eq:BRtautau} \\
{\rm A}_{\tau\tau} &=& (0.434 - 0.042 R_7) / (10^7 \,{\rm BR}_{\tau\tau}) \label{eq:Atautau} \,.
\end{eqnarray}

We have checked the sensitivity of these predictions to the various input parameters and to the scale $\mu_b$. The normalizations of the ${\rm BR}_{\ell\ell}$ can vary by up to ${\cal O}(10\%)$, but the functional dependence on $R_7$ varies only at the percent level. Thus, SM NLO calculations and more precise measurements of $m_t$ and $\alpha_s(m_Z)$ can significantly sharpen the predictions. There is very little sensitivity to $R_8$ (which nonetheless can be included). The normalizations of ${\rm A}_{ee}$ and ${\rm A}_{\mu\mu}$ are even more sensitive than the branching ratios to $\mu_b$, but a SM NLO calculation can reduce the uncertainties to less than $\sim10\%$. The $\tau^+\tau^-$ results are not very sensitive to those input parameters. And nonperturbative corrections are also quite small once the cuts on $\hat s$ are imposed.

Presently there exist \cite{ref:bsllexp} only some upper bounds on ${\rm BR}_{ee}$ and ${\rm BR}_{\mu\mu}$, ${\cal O}({\rm few}\times10^{-5})$, or about an order of magnitude above the SM. However, an increase of two to four orders of magnitude is expected \cite{ref:bsllfut} in the near future at the Tevatron. With such statistics we can look forward to rather precise measurements of most of the branching ratios and asymmetries.

Finally, we consider the anomalous magnetic moment of the muon. As shown by previous authors \cite{ref:gmtwo}, the high precision of calculations and measurements of $(g_\mu - 2)/2 \equiv a_\mu$ makes their comparison a promising harbinger of SUSY --- particularly when $\tan\beta$ is large. In the last of those papers, minimal gauge mediation was also considered, but without the predictions for $\mu$ and $\tan\beta$. We have recalculated the superpartner contributions to $a_\mu$, and are in agreement with the magnitude and sign of the results of Moroi in Ref.~\cite{ref:gmtwo} (after accounting for the erratum, and noting that the MGM predicts $\mu>0$ in that paper's conventions).  We find that once again the chargino amplitudes dominate the neutralino ones --- the latter never exceed 10\% of the former --- while within the chargino contribution the left-right amplitude completely dominates. Making only the approximation $v_D\ll v_U$, it reads:
\begin{equation}
a_\mu^{\tilde h^-} \simeq {3\alpha_2\over4\pi}\tan\beta {m_\mu^2 \mu M_2 F_\mu(x_{\tilde h_1 \tilde\nu},x_{\tilde h_2 \tilde\nu}) \over m_{\tilde \nu}^2 (M_{\tilde h_1^-}^2 - M_{\tilde h_2^-}^2)} 
\label{eq:Agmu}
\end{equation}
where $F_\mu(x_1,x_2) = f_\mu(x_1) - f_\mu(x_2)$, $f_\mu(x) = (3 - 4 x + x^2 + 2 \ln x)/(3 (1 - x)^3)$ and $x_{\tilde h_i \tilde\nu} = M_{\tilde h_i^-}^2/m_{\tilde \nu}^2$.  For our numerical results we use the complete  expressions.

The superpartner contributions must be added to the SM amplitudes before comparing with experiment. Combining the results of Refs.~\cite{ref:gmtwoSM} yields:
\begin{equation}
10^{10} a_\mu^{\rm SM} - 11\,659\,000 = 186 \pm 16
\label{eq:amuSM}
\end{equation}
while the current average experimental value is \cite{ref:gmtwoexp}:
\begin{equation}
10^{10} a_\mu^{\rm expt} - 11\,659\,000 = 230 \pm 84\,.
\label{eq:amuexpt}
\end{equation}
Evidently, the SM prediction is somewhat below the current experimental finding, although the discrepancy is less than a single {\it present} experimental standard deviation. We have carefully checked that the MGM generates a positive contribution to $a_\mu$, raising it above the SM prediction.

To calculate the predictions of the MGM for the various processes above, we use the results of Ref.~\cite{ref:rs}. All the supersymmetric standard model parameters are predicted in terms of the effective SUSY-breaking scale $\Lambda$ -- or equivalently the wino mass $M_2$ -- and the logarithm of the messenger scale $M_M$. There is only a slight sensitivity to the doublet and triplet messenger masses; we will assume $M_{M_3} = 1.3 M_{M_2}$. We consider messenger scales between a low $M_M \simeq 2 \Lambda$, which yields a large $\tan\beta = {\cal O}(50)$, and a rather high $M_M \simeq 10^4 \Lambda$, at which the fortuitous cancellation in $B$ is diminished so $\tan\beta = {\cal O}(25)$.  Much heavier messengers tend to generate instabilities in the scalar potential and generally face cosmological difficulties \cite{ref:rs}. With a low $M_M$ the one-step solution of the RG equations in Ref.~\cite{ref:rs} is accurate enough to extract all the parameters to the same order as the threshold corrections. A high $M_M$ reduces the accuracy of the one-step solution, but on the other hand the cancellations in $B$ and in the various amplitudes for $b\to s\gamma$ are reduced and with them the need for the high precision. Considering the inaccuracies in the one-step solution, in the leading-order calculation of $A_t$, and in the absence of threshold corrections to $\mu$ \cite{ref:rs}, we estimate that the uncertainties in our predictions for ${\rm BR}_\gamma$ and $a_\mu$ are comparable to the corresponding SM ones.

Our results are presented in Fig.~1. The predictions of the MGM for ${\rm BR}_\gamma$ ($\times 10^4$) and $a_\mu$ ($\times 10^{10}$) are indicated by the heavy red\  curves for five values of the messenger scale $M_M$. The curves become dashed when $M_2$ is light enough that the vacuum stability bounds of Ref.~\cite{ref:rs} are violated, primarily because the right-handed stau develops a charge-breaking VEV. Along each curve we indicate values of $M_2$, which increase as the MGM predictions tend towards those of the SM, shown as a circled dot. The SM uncertainty ellipse shown in gray should be applied to each of these predictions. The current experimental values are also indicated by a circled dot, surrounded by a large green\  $1\,\sigma$ ellipse and a larger, lighter green\  $95\%$ CL distorted ellipse. Under the horizontal ${\rm BR}_\gamma$ axis we present a corresponding axis of $R_7$ values, setting $R_8 \simeq 0$ as explained above. We then use $R_7$ to predict the branching ratios ($\times 10^{7}$) and asymmetries for $b\to s\ell^+\ell^-$ and show the results on the axes below, for those quantities sensitive to $R_7$. Note that the experimental and theoretical uncertainties shown for $b\to s\gamma$ yield allowed ranges for $R_7$ and hence for the $b\to s\ell^+\ell^-$ quantities, but the latter have in addition their own independent uncertainties. Thus the ranges shown in Fig.~1 indicate our ability to predict those quantities {\it after} the SM uncertainties associated with each process are significantly reduced. 

\begin{figure}[tb]
\centering
\leavevmode
\epsfxsize=11cm \epsfbox[20 50 527 645]{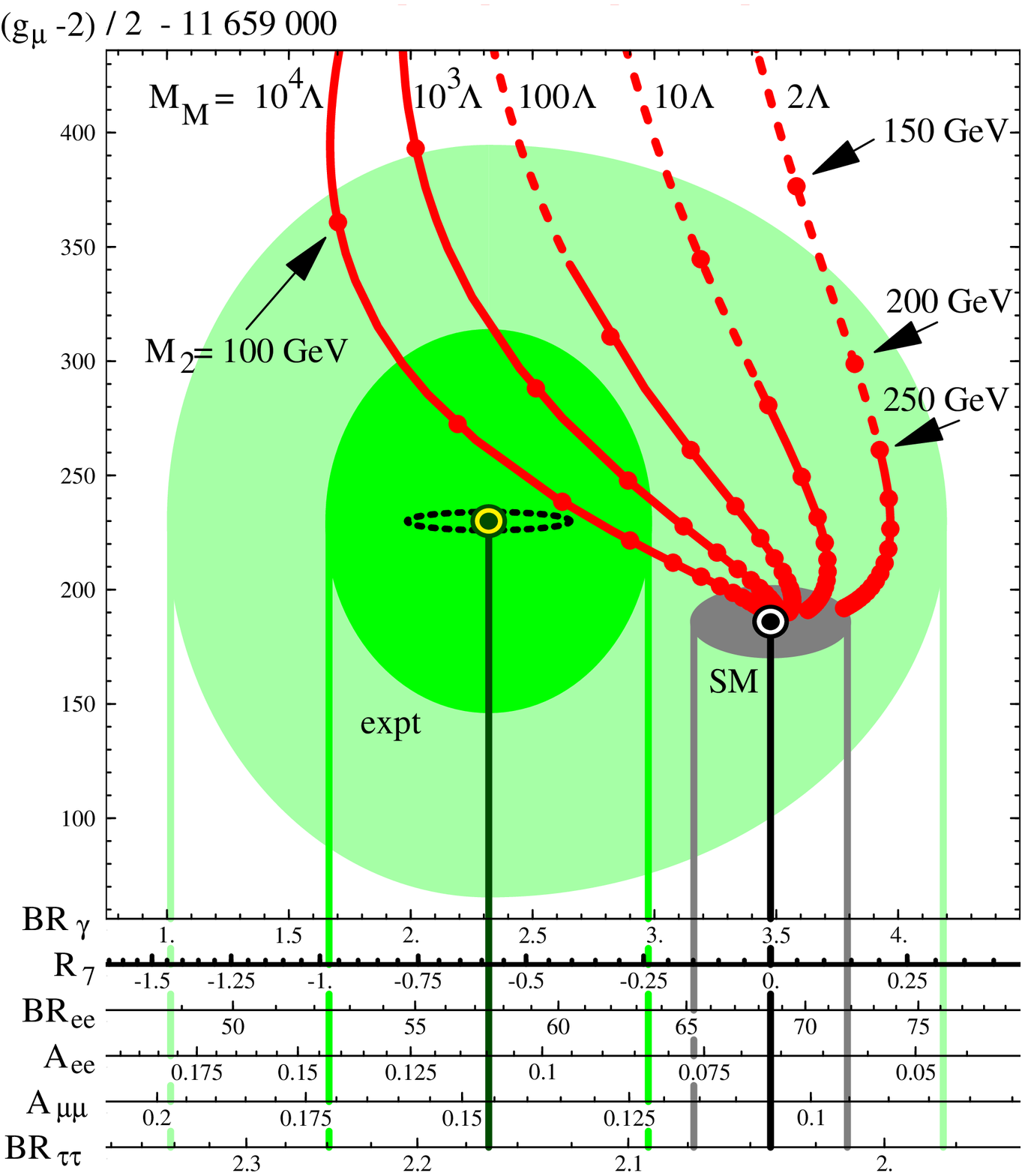}
\begin{quote}
{\small
Fig.~1. The MGM predictions (as heavy red\  curves) of ${\rm BR}_\gamma$ ($\times 10^4$) and $a_\mu$ ($\times 10^{10}$), compared to current and future (dashed) experiments. The extra horizontal axes are the correlated predictions for $R_7$ and the $B \to X_s\ell^+\ell^-$ branching ratios ($\times 10^{7}$) and asymmetries sensitive to $R_7$.}
\end{quote}
\end{figure}

We learn from Fig.~1 that with light messengers the MGM has little effect on ${\rm BR}\gamma$ (due to a precise cancellation between the charged-Higgs and superpartner amplitudes) but can somewhat raise $g_\mu-2$ in the experimentally-favored direction; the bounds imposed by $g_\mu-2$ measurements are not as strong as those due to vacuum stability. With a higher messenger scale, on the other hand, the MGM decreases ${\rm BR}\gamma$ while increasing $g_\mu-2$ beyond the SM and towards the current experimental values (as long as the wino is not extremely light). This is an encouraging sign for the MGM, although the current precision is inadequate to draw strong conclusions. However, help is on the way: CLEO data now being analyzed will sharpen ${\rm BR}_\gamma^{\rm expt}$, perhaps by a factor of two, while the E821 $g_\mu - 2$ experiment now operational at Brookhaven will yield \cite{ref:gmtwofut} a 20-fold improvement in precision. The corresponding near-future uncertainties appear as a dashed ellipse, centered (for lack of prescience) on the present experimental central values. The new data will dramatically constrain the likely scale of the superpartners and of the messenger fields. And since the theoretical errors in the $b\to s\ell^+\ell^-$ branching ratios and asymmetries could be reduced by NLO calculations similar to those already carried out for $b\to s\gamma$, the MGM would then precisely predict deviations from the SM of up to ${\cal O}(1)$ in ${\rm A}_{ee}$ and ${\rm A}_{\mu\mu}$, as well as smaller deviations in ${\rm BR}_{ee}$ and ${\rm BR}_{\tau\tau}$ --- at least some of which could be tested at hadronic colliders in the next few years.

We gratefully acknowledge discussions with I.~Bigi, T.~Moroi and C.~Wagner.


\begin{thebibliography}{99}

\bibitem[*]{EGmail}Electronic address:
emidio.gabrielli@roma1.infn.it

\bibitem[\dag]{USmail}Electronic address:
sarid@particle.phys.nd.edu

\bibitem{ref:GM}
M. Dine and A. Nelson, Phys. Rev. D {\bf 48}, 1277
(1993); M. Dine, A. Nelson and Y. Shirman, Phys. Rev. D {\bf 51},
1362 (1995); M. Dine, A. Nelson, Y. Nir and Y. Shirman, Phys.\ Rev.\ D
{\bf 53}, 2658 (1996); M. Dine, Y. Nir and Y. Shirman, Phys.\ Rev.\ D {\bf 55}, 1501 (1997) ; K.S. Babu, C. Kolda and F. Wilczek, Phys. Rev.
Lett. {\bf 77}, 3070 (1996).

\bibitem{ref:rs}
R. Rattazzi and U. Sarid, hep-ph/9612464.

\bibitem{ref:gslong}
E. Gabrielli and U. Sarid, in preparation.

\bibitem{ref:bsg}
R. Garisto and J.N. Ng, Phys.\ Lett.\ B {\bf 315}, 372 (1993); L.J. Hall, R. Rattazzi and U. Sarid, Phys.\ Rev.\ D {\bf 50}, 7048 (1994); R. Rattazzi and U. Sarid, Phys.\ Rev.\ D {\bf 53}, 1553 (1996); M.A. Diaz, Phys.\ Lett.\ B {\bf 322}, 207 (1994); F.M. Borzumati, Z. Phys. C {\bf 63}, 291 (1994); N.G. Deshpande, B. Dutta and S. Oh, Phys.\ Rev.\ D {\bf 56}, 519 (1997); S. Dimopoulos, S. Thomas and J.D. Wells, Nucl.\ Phys.\ {\bf B488}, 39 (1997); F.M. Borzumati, hep-ph/9702307.

\bibitem{ref:bbmr}
S. Bertolini, F. Borzumati, A. Masiero and G. Ridolfi,
Nucl.\ Phys.\  {\bf 353}, 591 (1991).

\bibitem{ref:bsgNP}
M.B. Voloshin, Phys.\ Lett.\ B {\bf 397}, 275 (1997);
A. Khodjamirian, R. Ruckl, G. Stoll and D. Wyler,
Phys.\ Lett.\ B {\bf 402}, 167 (1997);
Z. Ligeti, L. Randall and M.B. Wise, Phys.\ Lett.\ B {\bf 402}, 178 (1997);
A.K. Grant, A.G. Morgan, S. Nussinov and R.D. Peccei, hep-ph/9702380;
G. Buchalla, G. Isidori and S.J. Rey, hep-ph/9705253.

\bibitem{ref:bsgNLO}
K. Chetyrkin, M. Misiak and M. Munz, Phys.\ Lett.\ B {\bf 400}, 206 (1997); A.J. Buras, A. Kwiatkowski and N. Pott, hep-ph/9707482; and references therein.

\bibitem{ref:bsgEXPT}
M.S. Alam {\it et al.} (CLEO Collaboration), Phys.\  Rev.\ Lett.\ {\bf 74}, 2885 (1995).

\bibitem{ref:bsll}
A.J. Buras and M. Munz,  Phys.\ Rev.\ D {\bf 52}, 186 (1995); P. Cho, M. Misiak and D. Wyler, Phys.\ Rev.\ D {\bf 54}, 3329 (1996).

\bibitem{ref:hewells}
J.L. Hewett and J.D. Wells, Phys.\ Rev.\ D {\bf 55}, 5549 (1997).

\bibitem{ref:bsllexp}
R. Balest {\it et al.} (CLEO Collaboration), (preliminary) CLEO CONF 97-15;
C. Anway--Wiese {\it et al.} (CDF Collaboration), Phys.\  Rev.\ Lett.\ {\bf 76}, 4675 (1996);
S. Abachi {\it et al.} (D\O\ Collaboration), FERMILAB-CONF-96-253-E, contributed to ICHEP 96, Warsaw, Poland;
C. Albajar {\it et al.} (UA1 Collaboration), Phys.\ Lett.\ B {\bf 262}, 163 (1991).

\bibitem{ref:bsllfut}
T. Liu and S. Pakvasa, preprint UH-511-867-97.

\bibitem{ref:gmtwo}
J.L.\ Lopez, D.V. Nanopoulos and X. Wang, Phys. Rev. D {\bf 49}, 366 (1994);
U. Chattopadhyay and P. Nath, Phys.\ Rev.\ D {\bf 53}, 1648 (1996);
T. Moroi, Phys.\ Rev.\ D {\bf 53}, 6565 (1996), and erratum (submitted);
M. Carena, G.F. Giudice and C.E.M. Wagner, Phys.\ Lett.\ B {\bf 390}, 234 (1997).

\bibitem{ref:gmtwoSM}
T. Kinoshita and W.J. Marciano, in {\it Quantum Electrodynamics}, ed.\
T. Kinoshita (World Scientific, Singapore, 1990), p. 419;
T. Kinoshita, Phys.\  Rev.\ Lett.\ {\bf 75}, 4728 (1995);
A. Czarnecki, B. Krause and W.J. Marciano, Phys.\  Rev.\ Lett.\ {\bf 76}, 3267 (1996);
S.Eidelman and F. Jegerlehner, Z.\ Phys.\ {\bf C67}, 585 (1995);
K. Adel and F.J. Yndurain, hep-ph/9509378;
T. Kinoshita, B. Nizic and Y. Okamoto, Phys.\ Rev.\ D {\bf 31}, 2108 (1985);
J. Bijnens, E. Pallante and J. Prades, Phys.\  Rev.\ Lett.\ {\bf 75}, 1447 (1995), and erratum, ibid., 3781 (1995).

\bibitem{ref:gmtwoexp}
Particle Data Group, L. Montanet {\it et al.} Phys.\ Rev.\ D {\bf 50}, 1171 (1994).

\bibitem{ref:gmtwofut}
B.L. Roberts et al., contributed to ICHEP 96, Warsaw, Poland.



\end{thebibliography}
\end{document}